\documentclass[aps,pra,twocolumn,showpacs]{revtex4}  
\usepackage{graphicx}  
\usepackage{dcolumn}   
\usepackage{bm}        
\usepackage{amssymb,color}   
\usepackage{textcomp}
\begin{document}

\title{A two-band Bose-Hubbard model for many-body resonant tunneling in the Wannier-Stark system}

\author{Carlos~A.~Parra-Murillo,$^1$ Javier~Madro\~nero,$^2$ and Sandro Wimberger$^1$}
\affiliation{$^1$Institut f\"ur Theoretische Physik and Center for Quantum Dynamics, Universit\"at Heidelberg, 69120 Heidelberg, Germany\\
$^2$Departamento de F\'isica, Universidad del Valle, Cali, Colombia.}


\begin{abstract}
We study an experimentally realizable paradigm of complex many-body quantum systems, a two-band Wannier-Stark model, for which diffusion
in Hilbert space as well as many-body Landau-Zener processes can be engineered. A cross-over between regular to quantum chaotic spectra
is found within the many-body avoided crossings at resonant tunneling conditions. The spectral properties are shown to determine the evolution
of states across a cascade of Landau-Zener events. We apply the obtained spectral information to study the non-equilibrium dynamics of our 
many-body system in different parameter regimes.
\end{abstract}

\pacs{03.65.Xp, 05.45.Mt, 71.35.Lk, 37.10.Jk}

\maketitle

\section{Introduction} 
The rapid development of high precision techniques for the experimental control of ultracold quantum gases offers a clean way to study static
and dynamical properties of interacting many-body lattice systems \cite{IblochREP2008-1,IblochREP2008,IblochNAT2011}. Of particular interest are realizations
of strongly correlated or complex quantum systems composed of many particles. Experiments controlling the populations of higher orbitals and bands in
periodic potentials are now at hand \cite{Early,IblochPRL2007}. This facilitates the study of many degree-of-freedom systems. Moreover, 
many-body quantum quenches, non-equilibrium dynamics and real-time controlled sweep dynamics can be realized in the experiments 
\cite{MGreiner2011,Arimondo2011,InsbruckArxiv2013,Bloch2013}. This offers very good advantages for a better understanding of the underlying diffusion processes
taking place not only in real space, but, more generally, in the Hilbert space \cite{CarlosThesis2013,Wilkinson1988,APolkovnikov2013}. In this latter context, aspects
of integrability of quantum systems are crucial for predictions on, for instance, their relaxation dynamics and further thermalization 
\cite{IblocPrl2011,Deutsch1991,Srednicki1994,FHaakePRL2012,RigolPRL2012,LFSantosPRL2012,APolkovnikov2010,GogolinPRL2011}.

In this paper we present a case study of a complex many-body system including two strongly coupled energy bands. As we sketch in Sec.~\ref{sec:1a} and
in the appendix, this can readily be realized experimentally with ultracold bosons in accelerated (or equivalently tilted) optical lattices \cite{Oliver,Oliver-1,Oliver-2}. 
Our system represents a paradigm for the many-body physics, in which the dynamics can be steered by the parameters (tilt, interaction strength, potential depth), thus implementing 
very different dynamics. While the mean-field transport (weakly interacting limit) and the single particle limit are well studied for one-dimensional Wannier-Stark
systems \cite{KorschPRep2002,Oliver,Oliver-1,WimbPRL2007}, even the simplest many body version, a one-band Bose-Hubbard model with(out) tilt, allows one to tune between regular and quantum chaotic evolutions 
\cite{kolEPL2004,kol68PRE2003,Wolf2011,TomadimPRL2007,TomadimPRL2008,BW2008}. In our problem, depending on the choice of parameters,
full complexity in the interband transport in Hilbert space can be reached by sweeping predetermined initial states over resonant tunneling regions. 
This is possible since the strongest interband coupling occurs at resonantly enhanced tunneling (RET) between energy bands 
\cite{WimbPRL2007,KorschPRep2002}. At resonant conditions, we find a clear crossover from regular to quantum chaotic spectral statistics as a function of
a few system parameters. The complexity in the energy spectrum determines the transport across the many avoided 
crossings at RET when the force becomes time-dependent. This not only generalizes results on the weakly interacting limit \cite{PloetzJPB2010,PloetzEPJD2011}, 
but relates to the largely open problem of many-body Landau-Zener processes in the presence of strong particle interactions \cite{Wilkinson1988,TomadimPRL2007,IblochNAT2010}. 
As direct applications we show how the spectral properties influence the diffusion in Hilbert space. We characterize different realistic scenarios
for which relaxation toward equilibrium and spectral localization on the one hand and diffusion on the other hand take place. This is done with the help of controlled sweeps
through the interband many-body resonant regimes.

This paper is organized as follows: in Sec.~\ref{sec:1}, we introduce our two-band Bose-Hubbard model and the numerical methods implemented
for the diagonalization. Its spectral properties are presented in Sec.~\ref{sec:2}, where the conditions for the emergence of chaos 
are discussed along with predictions for the dynamics. In Sec.~\ref{sec:3}, we study the diffusion processes for different spectra and initial conditions
when driving the system through the resonant regime. Finally Sec.~\ref{sec:4} concludes the paper with a discussion of experimental ramifications.

\section{Many-Body Wannier-Stark Problem} \label{sec:1}
\subsection{The Two-band Model}\label{sec:1a}
Our Wannier-Stark system consists of ultracold bosonic atoms in a one-dimensional optical lattice. An additional Stark force
stimulates the quantum transport along the lattice \cite{Oliver,Oliver-1,Oliver-2,kol68PRE2003} and, at the same time, couples
the two lowest Bloch bands. The system we have in mind, see Eq.~(\ref{eq:01}), could be realized experimentally with ultracold bosons in a doubly periodic optical
lattice. For a convenient choice of the parameters, a well isolated two-band system can be engineered, thus neglecting the effects of the third and higher excited
Bloch bands (see Fig.~\ref{fig:apx01} in the appendix). Further details on the realization can be found in the appendix.

The corresponding many particle problem can be described in the tight-binding limit by a two-band Bose-Hubbard Hamiltonian 
\begin{equation}\label{eq:01}
  \hat{H}=\sum_{\beta=a,b}\hat{H}_{\beta}+\hat{H}_{1}+\hat{H}_{2},
\end{equation}
\noindent with the terms in Eq.~(\ref{eq:01}) defined by 
\begin{eqnarray}\label{eq:02}	
 \hat{H}_{\beta} &=&\sum^L_{l=1}-\frac{J_{\beta}}{2}\left(\hat{\beta}^{\dagger}_{l+1}\hat{\beta}_{l}+h.c.\right)+
 \frac{W_{\beta}}{2}\hat{\beta}^{\dagger 2}_{l}\hat{\beta}^2_{l}+ \varepsilon^{\beta}_l\hat{n}^{\beta}_l\,,\nonumber\\
 \hat{H}_1&=&\sum^L_{l=1}\sum_{\mu} \omega_B C_{\mu} (\hat{a}^{\dagger}_{l+\mu}\hat{b}_l+h.c.)\,,\nonumber\\
 \hat{H}_2 &=&\sum^L_{l=1}2W_x\hat{n}^a_l\hat{n}^b_l+
 \frac{W_x}{2}\left(\hat{b}^{\dagger}_{l}\hat{b}^{\dagger}_{l}\hat{a}_{l}\hat{a}_{l}+h.c.\right)\,.
\end{eqnarray}
The bosonic annihilation (creation) operators at the $l$-th site are $\hat{\beta}_l(\hat{\beta}^{\dagger}_l)$, and the number operators 
are $\hat{n}^{\beta}_l={\hat{\beta}}^{\dagger}_l{\hat{\beta}}_l$. $\beta$ is the band index, i.e., $\beta=a$ for the lower band and $\beta=b$
for the upper one. The on-site energies are given by $\varepsilon_l^{\beta}=\omega_B l+\Delta_g\delta_{\beta,b}$, the Bloch frequency is
~$\omega_B=2\pi F$ and $\Delta_g$ is the energy separation between the Bloch bands. $J_{\beta}$ are the hopping amplitudes. The on-site 
interparticle interaction in the bands is assumed to be repulsive with strength $W_{\beta}>0$. 
The coupling between the bands is given by:~$(i)$ the \emph{dipole-like} terms in $\hat H_1$ with strength proportional to $C_{\mu}$, 
where the integer index $\mu$ is symmetric around 0, and
 $(ii)$ the interaction terms with strength $W_x$ in $\hat H_2$. 

In the single particle picture, the interband coupling is maximal at specific tilts $F_r\approx \Delta_g/2\pi r$. At those values resonantly enhanced tunneling
(RET) occurs between levels located at wells separated by a distance $r$. This integer $r=l_{a}-l_{b}$ is called the order of the resonance \cite{WimbPRL2007}. 
The above resonance formula is modified to 
\begin{equation}\label{eq:03}
F_r=\Delta_g/2\pi\sqrt{r^2-4C_0^2} 
\end{equation}

\noindent by taking into account the Stark shift of the levels \cite{PloetzJPB2010}.

\subsection{The Floquet-Bloch operator and its numerical diagonalization}\label{sec:1b}
It is convenient to transform the Hamiltonian (\ref{eq:01}) into the interaction picture with respect to the external force, which removes the tilt $\sum_{l,\beta}\omega_Bl\hat{n}^{\beta}_l$ 
and transforms the hopping terms as: $\hat{\beta}^{\dagger}_{l+1}\hat{\beta}_l\rightarrow\hat{\beta}^{\dagger}_{l+1}\hat{\beta}_l\exp{(-i\omega_Bt)}$. In addition, 
in this procedure the dipole-like couplings with $|\mu|>0$ are transformed as: $\hat{a}^{\dagger}_{l+\mu}\hat{b}_l\rightarrow\hat{a}^{\dagger}_{l+\mu}
\hat{b}_l\exp{(-i\omega_B \mu Ft)}$. The gauge-transformed Hamiltonian is now translationally invariant and time-dependent with the fundamental 
period $T_B=2\pi/\omega_B$, the Bloch period, i.e. $\hat{H}(t+T_B)=\hat{H}(t)$. This condition holds because the remaining frequencies are integer
multiples of $\omega_B$. We are now allowed to impose periodic boundary conditions in space, i.e. by identifying $\hat{\beta}^{\dagger}_{L+1}=\hat{\beta}^{\dagger}_{1}$. Therefore, a
suitable basis for numerical diagonalization is given by the translationally invariant Fock states $\{|\gamma\rangle\}$ defined in Refs.~\cite{kol68PRE2003,TomadimPRL2007,TomadimPRL2008}. 
We can also work with the Floquet Hamiltonian $\hat{H}_f=\hat{H}(t)-i\partial_t$ \cite{Shirley1965}, for which the eigenvalue equation reads 
\begin{eqnarray}\label{eq:04}
\varepsilon_i\hat{1}\arrowvert \phi^{k}_{\varepsilon_i}\rangle&=&
\left({\hat H_0}- \omega_B k\hat{1}\right)\arrowvert \phi^{k}_{\varepsilon_i}\rangle+{\hat J}\arrowvert \phi^{k-1}_{\varepsilon_i}\rangle
+{\hat J}^{\dagger}\arrowvert \phi^{k+1}_{\varepsilon_i}\rangle\nonumber\\
&+&\sum_{\mu}\left[{\hat C}_{\mu}\arrowvert \phi^{k-\mu}_{\varepsilon_i}\rangle
+{\hat C}_{\mu}^{\dagger}\arrowvert \phi^{k+\mu}_{\varepsilon_i}\rangle\right].
\end{eqnarray}
\noindent Here we used multi-mode Fourier decomposition of the eigenstates of $\hat{H}_f$ \cite{THoShu1985}, i.e., $|\phi_n(t)\rangle=\sum_{k}\exp{(-ik\omega_Bt)}| \phi^{k}_{\varepsilon_n}\rangle$,
with $k=k_1+2k_2+...+(L-1)k_{L-1}$. The operator $\hat{H}_0$ contains all the time-independent terms of the gauge-transformed Hamiltonian $\hat{H}(t)$ and
the operator $\hat{J}$ and $\hat C_{\mu}$ are defined by the hopping and dipole-like transition terms $\hat{J}=-\sum_{l,\beta}J_{\beta}\hat{\beta}^{\dagger}_{l+1}\hat{\beta}_{l}/2$ and
$\hat C_{\mu}=\omega_B C_{\mu}\sum_la^\dagger_{l+\mu}b_l$, respectively. In order to diagonalize (\ref{eq:04}) we use the expansion 
$|\phi^{k}_n\rangle=\sum\nolimits_kA_{k,\gamma}|\gamma\rangle$, which implies that the Floquet operator is represented by a block matrix. Since $|C_{\mu}|$ drops
faster to zero as $|\mu|$ increases because of the decreasing overlapping between the Wannier states at different lattice sites, we can neglect all processes with $C_{|\mu|>r}$. 
In this paper, we restrict to resonances of order $r=1$ and $r=2$, then the Floquet matrix is reduced to a four block diagonal matrix (see appendix
2 \cite{CarlosThesis2013}) with every block size given by the dimension of the Hamiltonian~(\ref{eq:01}).

In order to compute the quasienergies $\varepsilon_i$ (eigenvalues of $\hat{H}_f$) we numerically diagonalized the Eq.~(\ref{eq:04}) by a Lanczos
algorithm \cite{Lanczos}. The quasienergies lie within the so-called Floquet zone (FZ): $\varepsilon_i\in [\varepsilon_0-\omega_B/2,\varepsilon_0+\omega_B/2]$ of width $\omega_B$ and 
centered at $\varepsilon_0$. We conveniently set $\varepsilon_0$ as a function of $F$ in order to improve the visualization of the spectrum in the different 
regions of interest. Due to the periodicity of the quasienergies the extended spectrum is given by the operation $\varepsilon_i\rightarrow\varepsilon_i+n_{fz}\omega_B$, with the index $n_{fz}$ of the FZ. 
For $N$ atoms distributed in $L$ lattice sites, the number of quasienergies is given by $\mathcal{N}_s=(N+2L-1)!/[LN!(2L-1)!]$, considering 
the reduction by a factor $L$ arising from the translational symmetry \cite{kol68PRE2003,TomadimPRL2007,TomadimPRL2008}. However, the effective dimension of 
$\hat{H}_f$ in Eq.~(\ref{eq:04}) is much larger: $\mathcal{N}_s \Delta k$, with $\Delta k = 10\ldots50$ being the number of Floquet components
needed to obtained a number $\mathcal{N}_s$ of convergent eigenstates. This latter procedure is equivalent to diagonalizing the evolution operator
integrated over one Bloch period $\hat{U}_{T_B}=\hat{\mathcal{T}}\exp\left[-i\int_0^{T_B} \hat{H}(t)dt\right]$, where $\hat{\mathcal{T}}$
is the time ordering operator. Nevertheless, the diagonalization of $\hat H_f$ has advantages with respect to the computation times for larger systems and
large $T_B$.

\section{Spectral properties of the two-band Wannier-Stark system}\label{sec:2}

\subsection{The single particle limit and the two-band manifold approach}\label{sec:2a}
\begin{figure}[t]
\centering 
\includegraphics[width=\columnwidth]{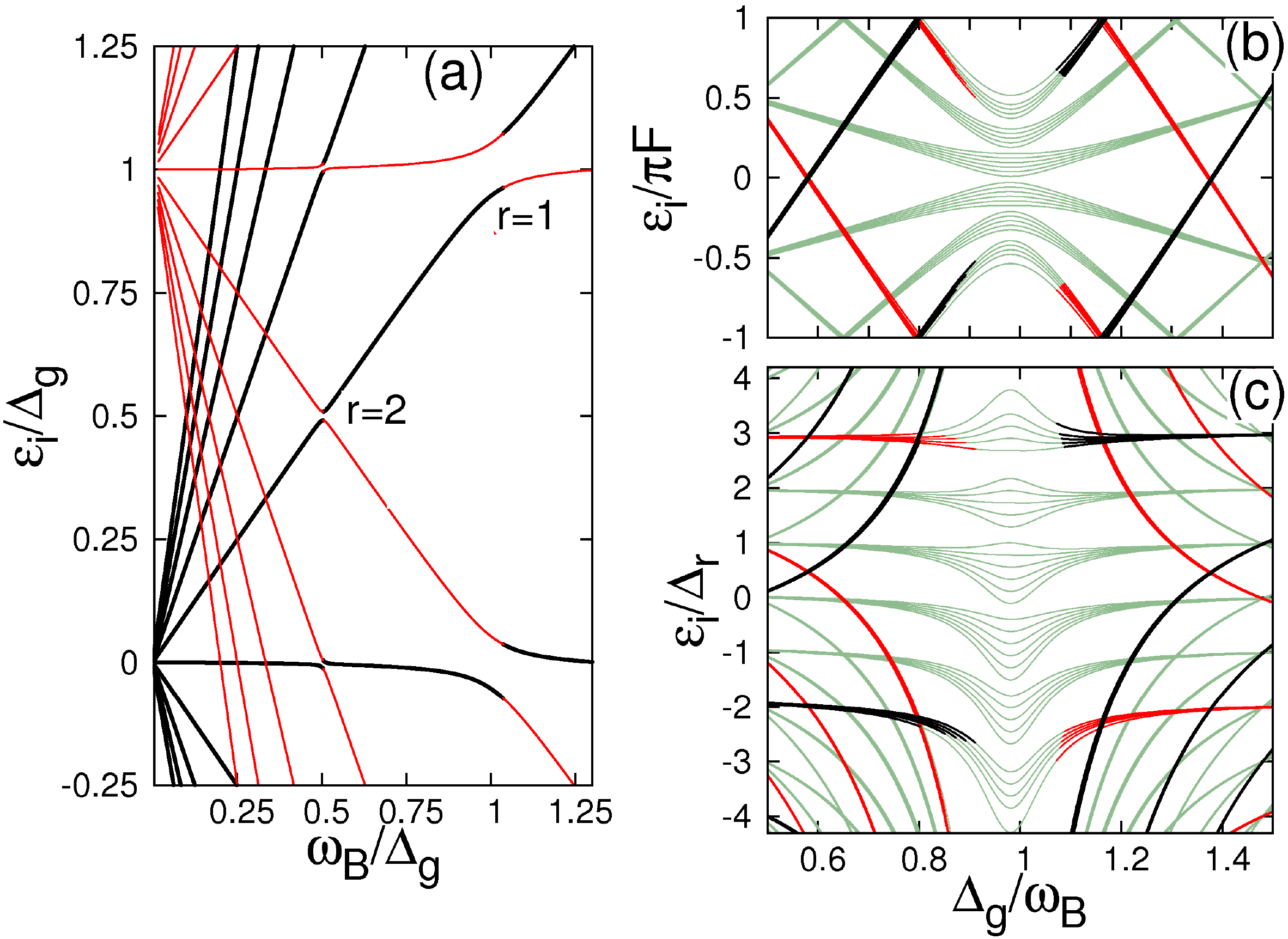}
\caption{\label{fig:1}(Color online)~The spectrum vs. Stark force $F$: (a) two-band Wannier-Stark ladders for the single particle case, for $\Delta_g = 0.796$ (corresponding to lattice parameters $V_0=5$ and $z_0=2.5$, see appendix). Avoided crossings appear at the resonances $F_{r}$, with width $\Delta^{\rm min}_r\ll\Delta_g$.~(b-c) The many-body spectrum for $N/L=5/3$
with no interparticle interaction, revealing the presence of the $M$-manifolds discussed in Sec.~\ref{sec:2}. The different lines correspond to eigenstates
of the type: lower-band-like states $\{|N\rangle_a\otimes|0\rangle_b\}$ (black lines), upper-band-like states $\{|0\rangle_a\otimes|N\rangle_b\}$ (thick red lines), and mixed-like states 
$\{|N-M\rangle_a\otimes|M\rangle_b\}$ (thin green lines). The remaining parameters are $C_0=-0.095$, $C_1=0.04$, $C_2=0.004$, $J_a=0.078$ and $J_b=-0.24$.}
\end{figure}

Let the force $F$ be the control parameter to analyze the spectrum in the plane $\varepsilon$-$F$ as shown in Fig.~\ref{fig:1}(a) for the single particle case. The 
gap $\Delta_g$, typically the largest energy scale in Eq.~(\ref{eq:01}) for experimental realization (see appendix), allows us to split up the spectrum into
equidistant subsets of states, each labeled by the upper band occupation number 
\begin{equation}\label{eq:05}
M\equiv\langle\varepsilon_i|\sum_l\hat{n}^b_l|\varepsilon_i\rangle. 
\end{equation}
 
In the off-resonant regime, where $F$ is not close to $F_r$, $M$ is a good quantum number since the eigenstates of $\hat H_f$ essentially correspond to specific basis states, i.e.,
to translationally invariant Fock states $|\gamma\rangle$. One can group these states into $N+1$ subsets of states with the same $M$ and dimension 
$\mathcal{N}_{M}=\frac{1}{L}{M+L-1\choose L-1}{N-M+L-1\choose L-1}<\mathcal{N}_s$. Hereafter we refer to those subsets as $M$-manifolds. 

In the noninteracting case, i.e. $W_{a,b,x}=0$, the internal manifold states are degenerate as shown by the level bunching in Fig.~\ref{fig:1}(b-c). 
The simplest case is that for $F=0$, for which the commutator $[\hat{H},\hat M]=0$. Then the Hamiltonian factorizes into a block matrix
$\hat{H}=\oplus_{M=0}^{N} \hat{\mathcal H}_M$. Note that the blocks $\hat\mathcal H_0$ and $\hat\mathcal H_N$ correspond to the independent Bose-Hubbard Hamiltonians
$\hat H_{\beta=a,b}$ respectively. Therefore, we can think of the Hamiltonian in (\ref{eq:01}) as two tilted Bose-Hubbard chains, connected through the mid-manifolds $0<M<N$ when $F\neq0$. 
The central manifolds contain all information about the interband coupling since they correspond to mixtures between states from both bands, for instance,
Fock states of type $|\vec{n}_{a,b}\rangle\leftrightarrow|N-M\rangle_{a}\otimes|M\rangle_{b}$. Furthermore, there is no direct coupling term between the blocks
$\hat\mathcal H_0$ and $\hat\mathcal H_N$. The interband coupling can be understood as the mixing of the $N+1$ manifolds, which is mainly induced by the one- and 
two-particle exchange terms in (\ref{eq:02}), that is $\hat H_1(F\neq0)$ and $\hat H_2(W_x\neq0)$. 

Around RET of order $r$, see Eq.~(\ref{eq:03}), the Hamiltonian (\ref{eq:01}) can be effectively transformed into a resonant Hamiltonian $\hat H_r$ by setting the reference system 
as $l_{a}=0$ with $l_{b}=-r$. Additionally, we define the manifold projectors 
\begin{equation}\label{eq:06}
\hat{P}_{M}=\sum\nolimits_{i}|\vec{n}_{a,b},M\rangle_i{}_i\langle\vec{n}_{a,b},M|, 
\end{equation}
where $|\vec{n}_{a,b},M\rangle\equiv|n^a_1,n^a_2,...\rangle\otimes|n^b_1,n^b_2,...\rangle$. The closure condition is given by $\sum_{M}\hat{P}_{M}=1$. This allows one to 
transform the Schr\"odinger equation $\hat H_r|\psi\rangle=E|\psi\rangle$ into the $M$ representation, where the resonant Hamiltonian becomes
\begin{equation}\label{eq:07}
  \hat{H}_r\simeq\sum^{N}_{M=0}\varepsilon^r_M|\psi_M\rangle\langle \psi_M|
  +\tilde{\omega}_B\left(|\psi_M\rangle\langle \psi_M+1|+h.c.\right).
\end{equation}

\noindent Here $\varepsilon^r_M=(\Delta_g-\omega_B r)M+(J_a-J_b)M$, $\tilde\omega_B\equiv\omega_B C_0\sqrt{M+1}$, $|\psi_M\rangle=\hat P_M|\psi\rangle$
and we used $N=N_a+N_b$, with $M\equiv N_b$. In this expression for $\varepsilon_M^r$, the order of the resonance (c.f. Sec.~\ref{sec:1a}) is approximated by $r\approx\Delta_g/\omega_B$. For typical parameters we have
that $\Delta_g,\omega_B\gg |J_b-J_a|$. Note that we disregard the dipole-like processes $|C_{|\mu|\geqslant 1}|$, which are only relevant at the exact resonance
inducing a splitting of the manifold levels. In this representation the Hamiltonian is clearly transformed into a tight-binding-type (TB) Hamiltonian for the manifolds,
where the first neighbor interaction is induced by a one-particle exchange with transition strength proportional to $\omega_B C_0$. Therefore, certain 
localization features are expected in energy space (as discussed in other contexts in~\cite{IzrailevPhysSkripta2001,APolkovnikov2013}), which in our case imply a high occupation probability of
a specific $M$-manifold. 

An important energy scale is given by the energy difference between neighboring manifolds $|\psi_M\rangle$ and $|\psi_M+1\rangle$,
which characterizes the one-particle exchange process (see Fig.~\ref{fig:2}(c)). This scale can be estimated by diagonalizing the $2\times 2$ Hamiltonian matrix
\begin{eqnarray}\label{eq:08}
H_{2\times 2} =
\left(
\begin{array}{cc}
\varepsilon^r_{M+1} & \omega_BC_0 \\
\omega_BC_0        & \varepsilon^r_M  \\
\end{array}
\right) \,,
\end{eqnarray}
\noindent from which we obtain 
\begin{eqnarray}\label{eq:09}
 \Delta_r=\Delta_g\sqrt{\left(1-\omega_B r/\Delta_g\right)^2+4(\omega_B C_0/\Delta_g)^2}.
\end{eqnarray}
\noindent The minimal width of the bow-tie-shaped many-body noninteracting spectrum in Fig.~\ref{fig:1}(b) is thus straightforwardly given by
$\Delta E=N\Delta^{\rm min}_r$, with $\Delta^{\rm min}_r=2\omega_B|C_0|$. 

\subsection{Interaction effects and manifold mixing at resonant tunneling}\label{sec:2b}
\begin{figure}[t]
\centering \includegraphics[width=0.99\columnwidth]{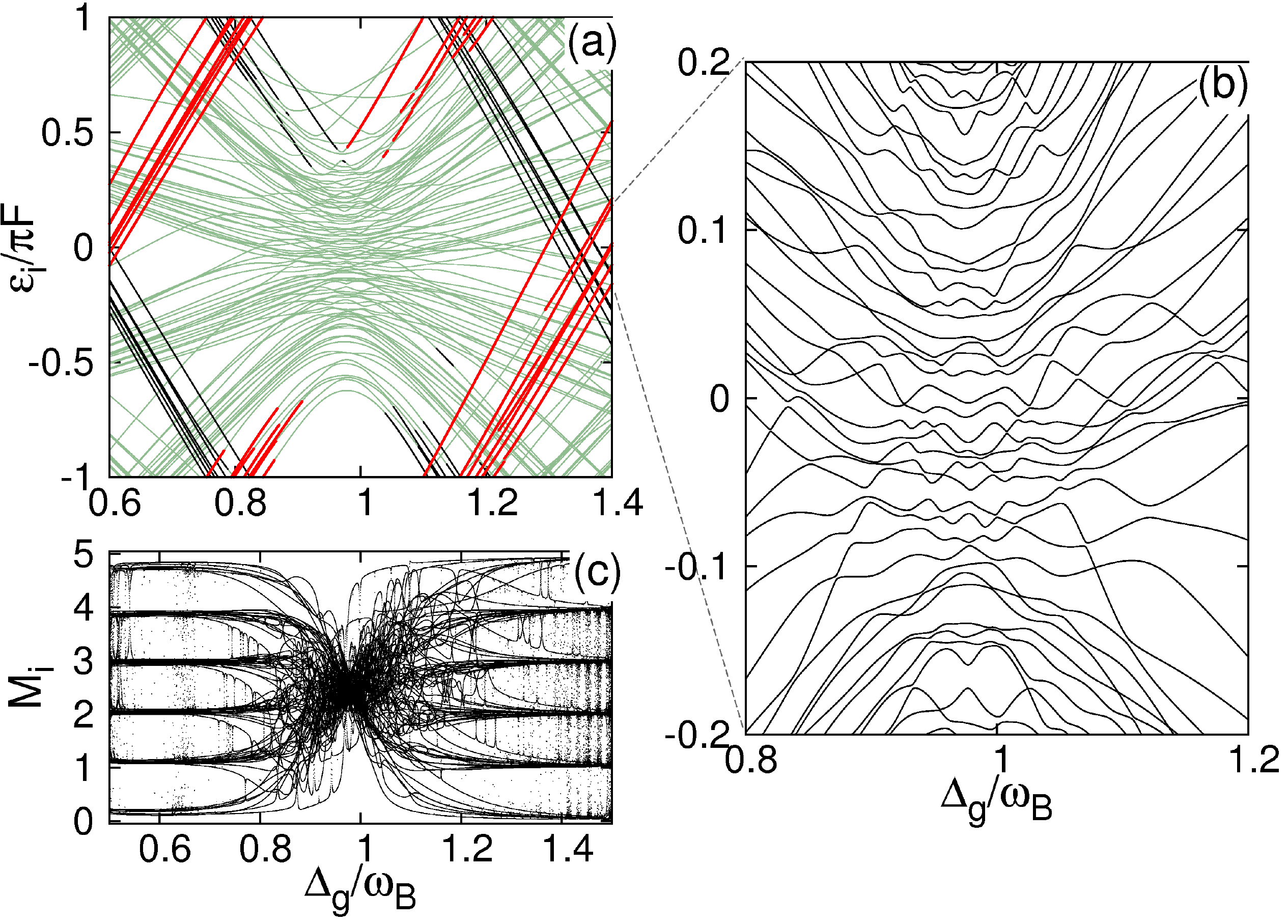}
\caption{\label{fig:2}(Color online)~Interaction effects: (a) Interacting many-body spectra for $N/L=5/3$ (see Fig.~\ref{fig:1}(a)). (b) Zoom
around the resonance position revealing the emerging cluster of avoided crossings. (c) Mani\-fold number $M_i$ for all Floquet eigenstates as a function
of the ratio $\Delta_g/\omega_B$. Here the mani\-fold structure is clearly seen before and after the single particle resonance $F_{r=1}=0.128$, characterized
by the bunches of eigenstates with approximately the same upper band occupation number $M$. The parameters are the same as those in Fig.~\ref{fig:1}, 
with additional interaction strengths $W_a=0.023$, $W_b=0.027$, $W_x=0.025$ (see Eq.~(\ref{eq:02})).}
\end{figure}

\begin{figure}[t]
\centering
\includegraphics[width=\columnwidth]{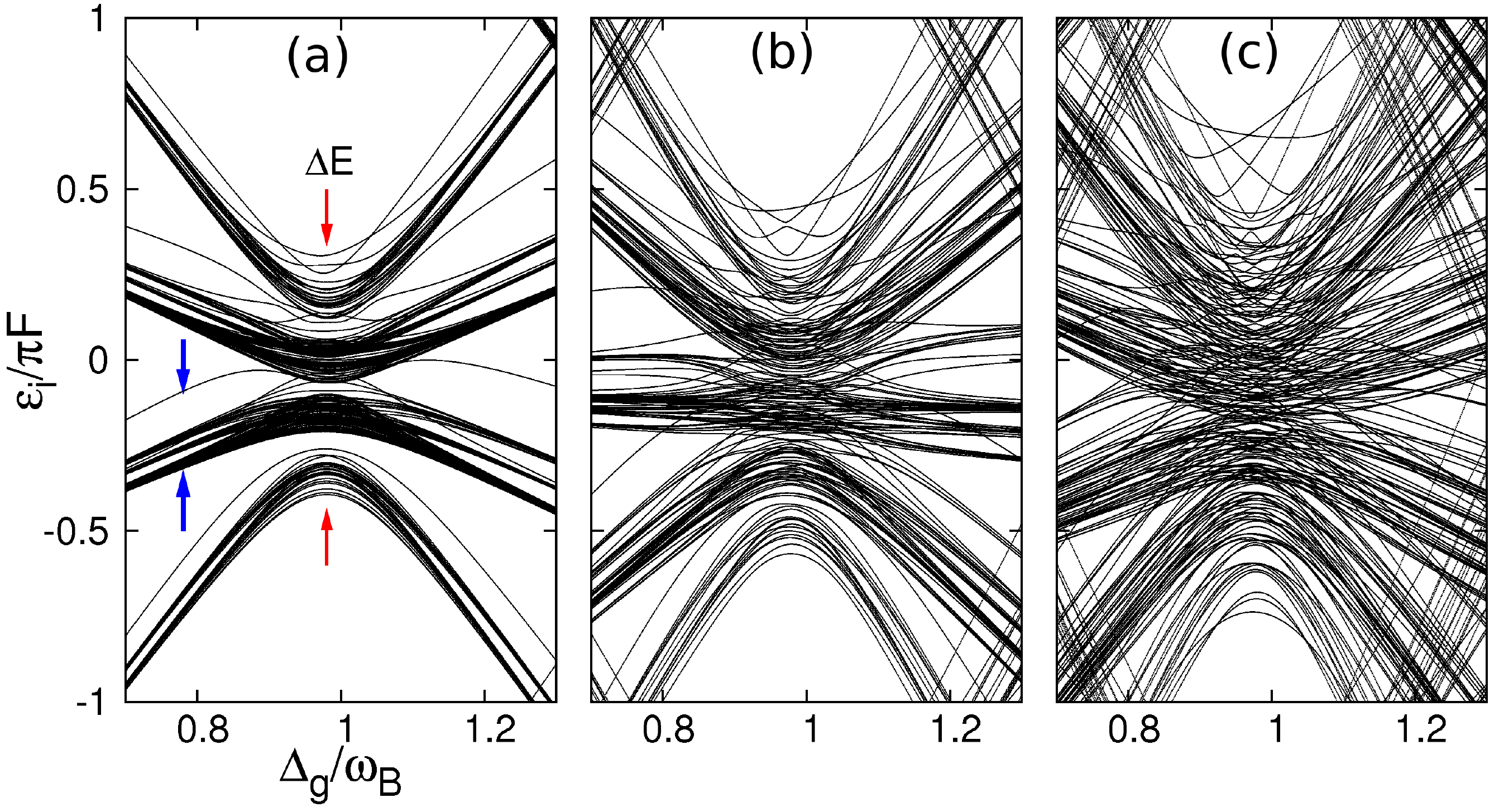}
\caption{\label{fig:3}(Color online)~Mani\-fold mixing: interacting many-body spectra as a function of the filling factor: (a) $N/L=3/13$, (b) $N/L=4/5$ and (c)
 $N/L=5/4$. Strong mani\-fold mixing occurs as $N/L$ increases due to the high density of avoided crossing in the resonant regime. The arrows in (a)
 represent: (red/thin) width of the many-body avoided crossing $\Delta E$ and (blue/thick) the maximal energy splitting of the central mani\-fold $M=1$. The parameters
 are the same as those in Fig.~\ref{fig:2}.}
\end{figure}

The interparticle interaction ($W_{a,b,x}\neq0$) splits up the internal manifold levels
and strong level mixing occurs at RET condition when the levels come closest. Then avoided crossings (ACs) appear due to the level repulsion, which 
arises from the lack of symmetries (see Fig.~\ref{fig:2}(b)). The number of ACs is the larger, the larger the filling factor $N/L$ (see Fig.~\ref{fig:3}). 
The maximal splittings by the on-site interparticle interaction occur due to those states with $M$ particles occupying a single particle level in one lattice
site (see blue arrows in Fig.~\ref{fig:3}(a)), for example $|N-M,0,..\rangle_a\otimes|M,0,...\rangle_b$. These are given by
\begin{eqnarray}\label{eq:10}
  (U^M_a)_{\rm max}&=&\frac{W_a}{2}(N-M)(N-M-1),\nonumber\\
  (U^M_b)_{\rm max}&=&\frac{W_b}{2}M(M-1),\nonumber\\
  (U^M_{ab})_{\rm max}&=&2W_x(N-M)M.
\end{eqnarray}

\noindent With these quantities we can compute the maximal mani\-fold splitting as $U(M)\equiv\max\{(U^M_{\beta=a,b})_{\rm max},(U^M_{ab})_{\rm max}\}$. 
Then we estimate the width $\Delta E$ of the many-body avoided crossing as
\begin{eqnarray}\label{eq:11}
\Delta E=N\Delta^{\rm min}_r+U(N)=N\Delta^{\rm min}_r+\frac{W_b}{2} N(N-1).
\end{eqnarray}
\noindent This follows from the fact that the maximal splitting is generated by those states with total particle number $N$ in one lattice site in the
upper Bloch band $\beta=b$. Both scales, $U(M=1)$ and $\Delta E$ are sketched by the two pairs of arrows in Fig.~\ref{fig:3}(a). 
The mixing in the spectrum is the strongest, the closer are the central mani\-folds, namely around $F_r$. Therefore, there are neither characteristic energy scales nor good quantum numbers. Conversely,
in the off-resonant regime the energy spectrum is characterized not only by the mani\-fold number $M$, but also by the numbers 
$\theta_{\beta}=\langle\varepsilon_i|\sum\nolimits_l\hat n^{\beta}_l(\hat n^{\beta}_l-1)/2|\varepsilon_i\rangle$ and 
$\theta_{x}=2\langle\varepsilon_i|\sum\nolimits_l\hat n^{a}_l\hat n^{b}_l|\varepsilon_i\rangle$. The latter numbers arise from the energy splitting 
induced by the interaction terms in Eq. (\ref{eq:02}). In this way, the eigenenergies can be approached by 
\begin{equation}\label{eq:12}
\varepsilon_i(M,\vec{\theta})\approx M_i\Delta_r+W_a\theta_{a,i}+W_b\theta_{b,i}+W_x\mu_{x,i}. 
\end{equation}
\begin{figure*}[t]
\centering \includegraphics[width=\textwidth]{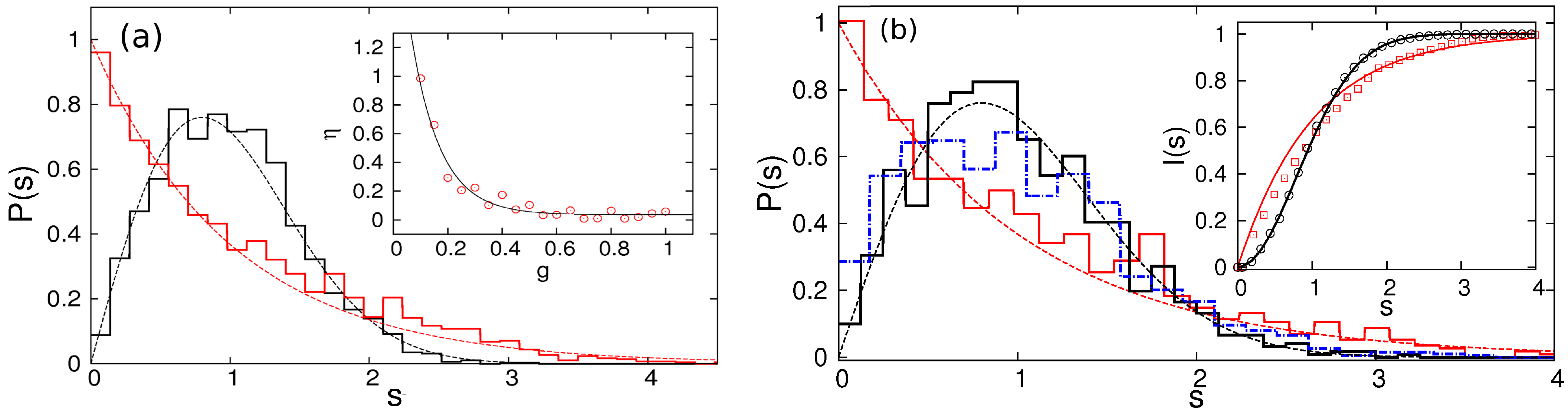}
\caption{\label{fig:4}(Color online)~Regular-to-chaotic transition: (a) The main panel shows the level spacing distribution $P(s)$ for $N/L=7/5$ ($\mathcal{N}_s=2288$)
for the interparticle control parameter (Sec.~\ref{sec:2}) $g=0.1$ (red/grey histogram, $\eta=0.98$) and $g=1.0$ (black histogram, $\eta=0.056$), and $W_x=W_a=W_b=0.025$.
The inset shows the parameter $\eta$ as a function of $g$, where the black line corresponds to an exponential fit. (b) The main panel depicts the level spacing
distribution $P(s)$ for three different filling factors: $N/L=3/25$ ($\mathcal{N}_s=848$, red/grey histogram), $N/L=4/11$ ($\mathcal{N}_s=1050$, blue dash-dotted
histogram) and $N/L=6/5$ ($\mathcal{N}_s=1001$, black thick histogram). The RMT distributions are those in dashed lines in both panels. The inset shows the
cumulative distribution $I(s)$ for the systems: $N/L=3/25$ (red/grey $\square$) and $N/L=6/5$ (black $\circ$). The solid lines represent the RMT prediction
for $I(s)$ Poisson (red/grey) and GOE (black). The other parameters are the same as in the previous figures.}
\end{figure*}

The effective region of mani\-fold mixing, the RET regime, is that for which the cluster of ACs is visible.~Therein, $M$, $\theta_{\beta}$
and $\theta_{x}$ are no longer good quantum numbers since even the identification of the otherwise most distant mani\-folds $M=0$ and $M=N$ becomes
difficult. Mani\-fold mixing is a local effect whenever $\Delta_g\gg J_{\beta},W_{a,b,x},C_s$ and $\Delta_g\approx r \omega_B$. Nevertheless,
global mixing can be engineered, for instance, by decreasing the energy band gap to a value comparable with the interaction strengths, i.e., $\Delta_g\approx W_{a,b,x}$.
Global mixing in the energy spectrum implies the destruction of the local resonances at $F=F_r$. Therefore, resonant tunneling generated by the interparticle 
interaction \cite{MGreiner2011,InsbruckArxiv2013} has the same relevance than the one generated by the interband coupling $W_{a,b,x}$.

In both cases local or global mani\-fold mixing, the spectral properties of (\ref{eq:01}) can be very complicated. Nevertheless, it is still possible to 
characterize the many-body spectrum in terms of the following subset of parameters: $(g,N/L,\Delta_g)$. 
Here $g$ is a prefactor that controls the strength of the interparticle interaction defined as $W_{a,b,x}\rightarrow gW_{a,b,x}$. Experimentally, changing $g$ 
is realized by varying the two-body scattering length via Feschbach resonances \cite{IblochREP2008,InsbruckArxiv2013}.

We now rescale the Hamiltonian (\ref{eq:01}) by the energy gap and then compute its commutator with the manifold number operator $\hat M$. This results in
\begin{eqnarray}\label{eq:13}
\left[\hat{H}/\Delta_g,\hat{M}\right]&=& \frac{1}{\Delta_g}([\hat H_1,\hat M]+[\hat H_2,\hat M])\nonumber\\
&=&\frac{\omega_B}{\Delta_g}\sum\nolimits_{l,\mu} C_{\mu}(\hat{a}^{\dagger}_{l+\mu}\hat{b}_{l}-h.c)\nonumber\\
&-&\frac{gW_x}{2\Delta_g}\sum\nolimits_l(\hat{b}^{\dagger}_{l}\hat{b}^{\dagger}_{l}\hat{a}_{l}\hat{a}_{l}-h.c).
\end{eqnarray}
\noindent From this equation we see that one- and two-particle exchange operators, corresponding to the two interband coupling processes, are mainly responsible for the mixing properties. Let us now fix the force to $F=F_r$ for which the interband coupling is maximized. Then we vary the filling factor, 
the band gap and the strength of the interparticle interaction. We have various cases: 
\begin{itemize}
\item[(i)] At resonance we have $\omega_B/\Delta_g\approx1/r$. If $g=1$, this implies that for high-order
resonances ($r>2$) $\hat H_2$ dominates only if $\Delta_g\lesssim  W_x$, otherwise the band coupling is 
weak and the commutator (\ref{eq:11}) goes to zero.
\item[(ii)] For $r=1$ and $\Delta_g\gg W_x$, $\hat H_1$ dominates, i.e., $\Delta_r$ is approximately a good energy scale. If $\Delta_g\sim W_x$
both $\hat H_1$ and $\hat H_2$ equally important.
\item[(iii)] At the condition $(ii)$, with $\Delta_g\sim W_x$, the filling factor plays an important role. In the case $N/L\ll1$, we are close to the single particle limit which is nearly
integrable. As the filling factor increases, so does the number of Fock states with double occupancies in a single lattice site, therefore there is a strong interplay between one- 
and two-particle exchanges. This naturally induces an enhancement of the manifold mixing. 
\end{itemize}

In terms of the $M$-manifolds, to consider the two-particle exchange process introduces a second neighbor transition term in our tight-binding
Hamiltonian of Eq.~(\ref{eq:07}). Such types of extended TB-type Hamiltonians are usually non-integrable (see refs. \cite{RigolPRL2012,LFSantosPRL2012} and references therein). 
We conclude that the many-body spectrum is strongly mixed when $\hat H_1$ and $\hat H_2$ have the same relevance, i.e., for the conditions
$|C_0|/r\sim W_x/2\Delta_g$ and $N/L\sim 1$. The latter can be achieved in both local, i.e. at RET, and global, i.e., for $\Delta_g\approx W_{a,b,x}$, manifold mixing.

\subsection{Emergence of many-body quantum chaos}\label{sec:2c}
We now investigate the many-body spectra by means of random matrix measures \cite{Haakebook}. We study the level spacing (or local gap) distribution
$P(s_i)$ with $s_i=\varepsilon_{i+1}-\varepsilon_i$, where $\langle s_i\rangle=1$, after an appropriated unfolding procedure \cite{CarlosThesis2013,Haakebook}. The crossover between 
regular (Poisson), $P_P(s)=\exp(-s)$, and quantum chaotic (Wigner-Dyson or GOE) statistics, $P_W(s)=\pi s\exp(-\pi s^2/4)/2$, can be reached in several ways. 

First, for an energy band gap $\Delta_g\lesssim 1$, we found that all systems with $N/L\sim 1$ are fully chaotic as shown in the main panel of
Fig.~\ref{fig:4}(a) for $N/L=7/5$. This is expected according to the commutator (\ref{eq:13}) and its respective discussion in the previous subsection. 

Secondly, for fixed filling factor, $N/L\sim1$, quantum chaos can be tuned by the prefactor $g$ of the interparticle interaction terms. In order to check this crossover,
we compute the parameter 
\begin{equation}\label{eq:14}
  \eta =\frac{\int^{s_0}_0\left(P(s)-P_W(s)\right)ds}{\int^{s_0}_0\left(P_P(s)-P_W(s)\right)ds},
\end{equation}
\noindent where $s_0=0.4729...$ is the intersection point between the distributions $P_P(s)$ and $P_W(s)$. $\eta$ is plotted as a function of $g$ in the 
inset of fig.~\ref{fig:4}(a). Herein $\eta=1$ for a perfect poissonian distribution and $\eta=0$ for a perfect the Wigner-Dyson distribution.

Deviations from the limiting random matrix distributions are found for filling factors approaching the single particle limit for $N/L\ll1$.
This is seen in Fig.~\ref{fig:4}(b) for $N/L=3/25$ and $N/L=4/11$.

Due to the high dimensional parameter space of our system perfect poissonian distributions are not easy to find at the RET domain.
On the other hand, good chaotic distributions are straightforwardly obtained. This is shown in the inset of
the Fig.~\ref{fig:4}(b) where the respective cumulative distribution is plotted, that is $I(s)=\int_0^sP(s')ds'$. For the remaining combinations of
parameters, we always obtain deviations characterized by non fully chaotic level spacing distributions \cite{CarlosThesis2013}. $N/L\sim1$ and
$|C_0|\sim W_x/2\Delta_g$ are similar conditions for the emergence of quantum chaos as in the one-band Bose-Hubbard models studied in 
\cite{kol68PRE2003,TomadimPRL2007,TomadimPRL2008}. Yet, in our two-band model, $F$ can also be large and the RET allows us to squeeze many-body energy levels in order to
enforce a chaotic level structure. Hence, our new model allows us to switch between more or less regular and chaotic regimes in the vicinity of $F_r$ (see the manifold
picture in Sec.~\ref{sec:2a} and Fig.~\ref{fig:3}).

As a final remark, we did not take into account, for our spectral analysis, those systems for which the greatest common divisor $\gcd(N,L)$ is larger
or equal to one, due to the existence of a temporal symmetry of the Hamiltonian as reported in Ref.~\cite{kol68PRE2003}. In the following,
we discuss important consequences that emerge from the spectral properties studied so far for the many-body Wannier-Stark system defined by Eq.~(\ref{eq:01}). 

\section{Diffusion in Hilbert Space}\label{sec:3}

\subsection{Eigenstate diffusion}\label{sec:3a}
The structure of avoided crossings presented above provides a perfect setup for studying dynamical processes generated by a cascade 
of single Landau-Zener (LZ) events around $F_r$. 

{\it Time evolution}: We now focus on the diffusion process triggered by the parametric time evolution of different initial conditions with 
$F(t)=F_0 + \alpha t$, and with $\alpha=\Delta F/\Delta T$. In analogy to the LZ problem \cite{LZ} we use a linear sweep.
Here $\Delta F$ represents the effective extension of the RET regime
and $\Delta T$ is the time needed to evolve the initial state from a starting tilt $F_0$ to the final one $F_f$ (see Fig.~\ref{fig:5}(a)), 
i.e., the sweeping time. A reasonable value for $\Delta T$, and hence for the sweeping rate $\alpha$, is given by the Heisenberg relation 
$\Delta Td \approx 1$, where $d\approx\Delta E/\mathcal{N}_s$ is mean level spacing of the many-body spectrum at $F_r$. We now rewrite
the Hamiltonian as follows
\begin{equation}\label{eq:15}
 \hat{H}(t) = \hat{H}_0+\hat{J}^{\dagger}e^{-i2\pi F(t) t} + \sum\nolimits_{\mu}\hat{C}^{\dagger}_{\mu}e^{-i2\pi \mu F(t) t} + h.c., 
\end{equation}
\noindent with $\hat H_0$, $\hat J$ and $\hat C_{\mu}$ as defined in Sec.~\ref{sec:1b}. We can study two types of dynamics using (\ref{eq:15}): first, 
by fixing the Stark force $F(t)=F$. This implies that the Hamiltonian is temporally periodic and fulfills all properties described Sec.~\ref{sec:1a}.
The time evolution of the initial state $|\psi(0)\rangle$ is thus obtained through stroboscopic quantum maps 
$|\psi(m+1)T_B\rangle=\hat U_{T_B}|\psi(mT_B)\rangle$, with $m$ an integer. Secondly, when considering the time-dependent pulse $F(t)=F_0+\alpha t$, the periodicity is broken. 
The temporal evolution must then be explicitly computed, e.g., by using a fourth-order Runge-Kutta method. In addition, $\hat H(F(t))$ does no longer preserve
the time-reversal symmetry, therefore the expansion coefficients of the state $|\psi(t)\rangle$ in any basis are in general complex numbers.

To determine the parameter regime for the dynamical evolution we define the parameter $\lambda\equiv\alpha/d\,\Delta F$. The diabatic passage (or sudden quench) is set
by $\lambda\gg1$. We expect an adiabatic evolution for $\lambda\ll1$, and a non-adiabatic one for $\lambda\sim 1$. Hereafter we concentrate on the 
non-adiabatic regime when driving the system through a single resonance (see Fig.~\ref{fig:5}(a)). Furthermore, we set the time scale to be the Bloch 
period defined by the tilt for the exact single-particle resonance $F_r$, that is, $T_B=1/F_r$. The Bloch period is small when $F_r$ is large. Therefore 
for practical implementations it is useful to concentrate on the dynamics across a first-order resonance. We have already seen in Sec.~\ref{sec:2c} that
the manifold mixing is enhanced because of the chaotic spectral properties of the RET regime for $N/L\sim 1$.

{\it Initial condition}: To study the emerging diffusion process in Hilbert space we have two natural choices for initial conditions: the Floquet eigenstates at any fixed force 
$|\varepsilon_i(F)\rangle$ and the translationally invariant Fock basis states $|\gamma\rangle$. The two set of states map one-to-one onto each other with 
probability $\gtrsim 80\%$ in the off-resonant regime, due to the presence of the $M$-manifolds. In this way if we choose $|\gamma\rangle$ as initial state, it 
is well localized in energy space (see Fig.~\ref{fig:2}(c)). In this sense, we have, without loss of generality, generic initial
conditions \cite{Deutsch1991,RigolPRL2012}. If the initial state is a Floquet eigenstate at $F_0$, then it is, by definition, well localized in the instantaneous 
spectrum at $F_0$. 

{\it Protocol}: $(i)$ The initial state $|\psi(0)\rangle$ is chosen to be, for instance, a Fock state with a well defined upper band occupation number $M$. This may
be prepared in the flat lattice condition, i.e., at $F=0$. $(ii)$ Then we evolve $|\psi(0)\rangle$ by suddenly ($\lambda\gg1$) ramping the lattice as:
$F=0\rightarrow F=F_0$. This allows us to set a non-equilibrium scenario, as sketched in Fig.~\ref{fig:5}(a). $(iii)$ Next, the state $|\psi(F_0)\rangle$
is non-adiabatically driven ($\lambda\sim1$) across the many-body AC from $F=F_0$ to $F=F_f$. When the evolution starts at $F_0$ a fast coupling of the initial state
with the local eigenstates is expected, since the spectrum at $F_0$ is highly mixed. Yet, if the spectrum at $F_0$ is well described in
terms of manifolds, $|\psi(F_0)\rangle$ mixes in a first instance with the eigenstates members of the same mani\-fold via hopping transitions, before 
it mixes states from other mani\-folds. This latter process gives rise to mani\-fold mixing in time, and hence, the diffusion in energy space. 

\subsection{Localization-delocalization transition}\label{sec:3b}

To quantify the diffusive processes across the ACs, we compute the probability amplitudes $C_i(t) \equiv \langle \varepsilon_i(F_k)|\psi(t)\rangle$, where $\{|\varepsilon_i(F_k)\rangle\}$ is the set of local
Floquet eigenstates at the instantaneous tilt $F(t=\Delta T_k)=F_k$, with $\Delta T_k=T_k-T_0$. As a function of the local energy space, the distribution of 
the probabilities $|C_i(t)|^2$ can be represented in terms of the \emph{local density of states} (LDOS) \cite{TDittrich1991,LFSantosPRL2012}:
\begin{equation}\label{eq:16}
P_{\psi}(\varepsilon,t)=\sum\nolimits_{i}|C_i(t)|^2\delta(\varepsilon-\varepsilon_i), 
\end{equation}
which allows for a visualization of the transit of the state $|\psi(t)\rangle$ across the ACs.~At $F_0$, 
$P_{\psi}(\varepsilon,T_0)$ is $\delta$-shaped (as indicated by the arrow in Fig.~\ref{fig:6}). As the tilt increases with the time, $P_{\psi}(\varepsilon,t)$ starts to delocalize
due to the multiple LZ transitions induced by the cluster of avoided crossings.~The diffusion depends on $\alpha$ \cite{Wilkinson1988}, but it is also highly sensitive to
the type of statistical distribution of the spectrum in the vicinity of $F_r$. In Fig.~\ref{fig:6}(a), we show the evolution of $P_{\psi}(\varepsilon,t)$ corresponding to $N/L=4/11$, with  
the initial state defined by a Floquet eigenstate with mani\-fold number $M= [N/2]$, with $[\cdots]$ standing for the integer part. For this system, the spectrum at $F_r$ presents deviations of the full
quantum chaotic regime as previously shown (see Fig.~\ref{fig:4}). The incoming state is well localized in energy space and its localization is preserved with high
probability after the passage through the RET regime, despite partial delocalization of $|\psi(t)\rangle$ around $F_r$. An initially localized state can stay 
well localized by two mechanisms: $(i)$ by a fast diabatic driving across the many-level AC, similarly to a diabatic crossing in a two-level Landau-Zener system;
$(ii)$ in the non-adiabatic dynamical regime the instantaneous state can exchange its character ($M$) with the local eigenstates during the crossing through the RET domain.
The outgoing state may be characterized either by the same mani\-fold number $M$ (see Fig.~\ref{fig:6}(a)) or by a different one. This latter implies
a change of the direction of the LDOS in course of time in the plane $\varepsilon-F(t)$. The exchange of character is inherited from a two-level AC 
\cite{Arimondo2011,CarlosThesis2013}. 
\begin{figure}[t]
\centering
\includegraphics[width=0.8\columnwidth]{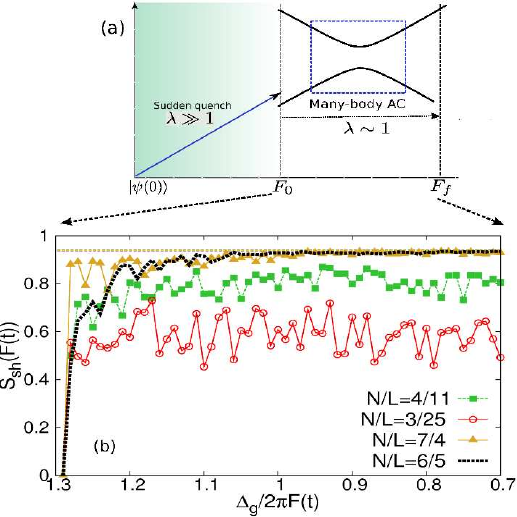}
\caption{\label{fig:5}(Color online) Diffusion: (a) Sketch of the sweeping process. The initial state $|\psi(0)\rangle$ is prepared at $F=0$ and
suddenly evolved by means of a quench from $F=0$ to $F=F_0$ ($\lambda\gg1$). After this process the state is then further evolved but non-adiabatically
($\lambda\sim 1$) from $F=F_0$ to $F=F_f$ across the many-body AC during a finite time $\Delta T$. The resonance order is $r=1$, with $F_{r=1}=0.045$. 
(b) Spreading of the evolved state $|\psi(t)\rangle$ measured by the Shannon entropy $S_{\rm sh}$ when crossing dynamically the RET regime.
The parameter are $\lambda=0.7$, $\Delta_g=0.285$, $J_a=0.0382$, $J_b=-0.0417$, $W_a=0.028$, $W_b=0.029$, $W_x=0.029$, $C_0=-0.096$, $C_1=0.046$, and 
$C_2=0.008$.}
\end{figure}

Localization properties of $P_{\psi}(\varepsilon,t)$ are analyzed by computing its second moment, from which we obtain 
\begin{equation}\label{eq:17}
\int \rho(\varepsilon) P^2_{\psi}(t)d\varepsilon =\sum_i|C_i(t)|^4\equiv\xi_{\psi}(t). 
\end{equation}
$\xi_{\psi}(t)$ is the so-called inverse participation ratio and $\rho(\varepsilon)$ is the density of states. In this way, the spreading over the
local instantaneous spectrum can be quantitatively characterized by the average inverse participation ratio \cite{TDittrich1991,ZelevinkyPRP1996}, and similarly by the
Shannon entropy \cite{kolEPL2004} both defined as
\begin{eqnarray}\label{eq:18}
\xi(F(t)) &\equiv& \left\langle\sum^{\mathcal{N}_s}_{i=1} |C_i(t)|^4\right\rangle_{\psi}, \\
S_{\rm sh}(F(t))&\equiv& \left\langle-\sum^{\mathcal{N}_s}_{i=1} 
\frac{|C_i(t)|^2}{\log_{10}\mathcal{N}_s}\log_{10}|C_i(t)|^2\right\rangle_{\psi}.
\end{eqnarray}
\noindent The average $\langle\cdot\rangle_{\psi}$ is taken over a large set of similar initial conditions $\{|\psi(0)\rangle\}$ with $M=[N/2]$.
The measures in (\ref{eq:16}) depend on the choice of basis to compute the coefficients $C_i$. In the case of complete delocalization the coefficients
$\{|C_i|\}$ fluctuate around the equipartition condition $|C_i|=1/\sqrt{\mathcal{N}_s}$. Therefore the localization measures (\ref{eq:16}) converge to
their respective minimal values, which can be computed under the assumption of complete randomness, i.e. no correlations between the coefficients. 
The set of coefficients $C_i$ satisfies a normalization condition $\sum_{i}|C_i(t)|^2=1$. Therefore we have $\mathcal{N}_s-1$ independent contributions. 
In the presence of chaos, the randomness of the above set of coefficients is guaranteed. Then by defining $y=|C_i|^2/\langle c^2\rangle$, with 
$\langle c^2\rangle$ being the average probability, the resulting distribution $f(y)$ follows a Porter-Thomas distribution 
\cite{Haakebook}. Because the coefficients $C_i$ are in general complex numbers due to the breaking of time-reversal symmetry of (\ref{eq:01}) when 
considering $F(t)$, we must use the GUE ensemble \cite{Haakebook}. For this ensemble RMT predicts  
$f(y)=\exp(-y)$. Following \cite{ZelevinkyPRP1996} and using $f(y)$, we compute the GUE (or statistical) limits of
the localization measures as 
\begin{eqnarray}\label{eq:19}
 \xi&=&\mathcal{N}_s\langle c^2\rangle^2\int^{\infty}_0dy\; f(y) y^2\\
 S_{\rm sh}&=&\mathcal{N}_s\int^{\infty}_0dy\; f(y) y\langle c^2\rangle \ln( y\langle c^2\rangle).
\end{eqnarray}

\noindent These integrals are accessible, from which we obtain:
\begin{equation}\label{eq:20}
\xi_{\rm gue}=\frac{2}{\mathcal{N}_s}\,,\;\; S_{\rm sh}^{\rm gue}=1-\frac{\sigma_c}{\ln(\mathcal{N}_s) }\,,
\end{equation}
\noindent with $\sigma_c=0.422784$ \cite{CarlosThesis2013}. We represented $S^{\rm gue}_{\rm sh}$ by the horizontal dashed lines in Fig.~\ref{fig:5}(b) along with 
the Shannon entropy $S_{\rm sh}(F(t))$ for different filling factors.

In case of deviations from the chaotic level spacing distributions ($N/L=3/25,4/11$), the time-evolved state does never reach the GUE 
limits, but remains localized instead (see Fig.~\ref{fig:5}(b)). The maximization of the entropy implies a dynamical equilibrium
\cite{LFSantosPRL2012,GogolinPRL2011}. Under this condition, the density operator $\hat\rho(t)=|\psi(t)\rangle\langle \psi(t)|$ diagonalizes in the local energy basis. We thus get 
$\langle\varepsilon_i|\hat\rho(t)|\varepsilon_j\rangle\approx|C_i|^2\delta_{i,j}$, since the off diagonal terms drop to zero. Then the Shannon and 
von Neumann entropies coincide \cite{APolkovnikov2010}. At this point, it is easily noticed that under chaotic conditions the density operator also 
diagonalizes in the Fock basis. This implies no further (re)localization in the course of the evolution, hence strong mixing of the complete 
set of mani\-folds is obtained. 
\begin{figure}[t]
\centering
\includegraphics[width=0.65\columnwidth]{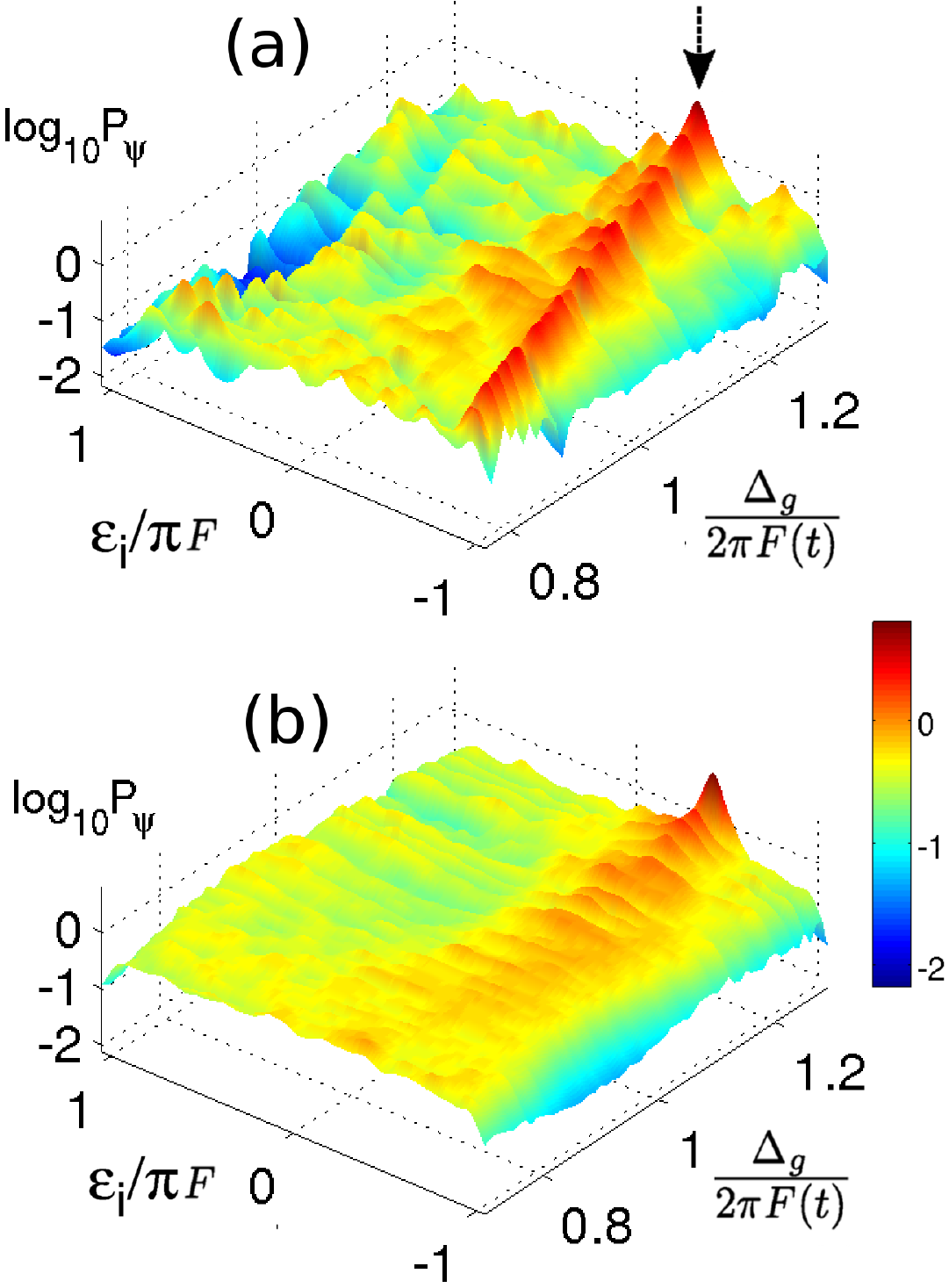}
\caption{\label{fig:6}(Color online) Diffusion: Temporal evolution of the initial state $|\psi(0)\rangle$ across the many-body AC 
represented by the local density of states $P_{\psi}(\varepsilon,t)$ as defined in Eq.~(\ref{eq:16}). The panels show the respective transit through
the energy spectra for (a) $N/L=4/11$ and (b) $N/L=6/5$. The remaining parameters are those of Fig.~\ref{fig:5}(b).}
\end{figure}

\subsection{Spectral ergodicity and relaxation toward equilibrium}\label{sec:3c}
We have seen previously that in the course of the time evolution, the system undergoes a dynamical diffusion (see Fig.~\ref{fig:5}(b)) 
in the accessible Hilbert space. This diffusive spreading is much stronger for chaotic spectra ($N/L=6/5,7/4$) than for poissonian or mixed ones 
($N/L=3/25,4/11$). The latter manifests itself in a (re)localization during the passage through the RET regime. The LDOS, locally in energy space,
 is thus described by the {\it Breit-Wigner} formula \cite{BreitWigner1936}:
\begin{equation}\label{eq:18-a}
P_{\psi}(\varepsilon,t)\sim\frac{1}{\pi}\frac{\Gamma^2/4}{(\varepsilon-\varepsilon_0)^2+\Gamma^2/4}, 
\end{equation}
\noindent where $\Gamma$ is width of the distribution and $\varepsilon_0$ its mean position in the spectrum. In the fully quantum chaotic (Wigner-Dyson distributed)
case such a (re)localization does not take place, as shown in Fig.~\ref{fig:5}(b) for $N/L\gtrsim 1$. $P_{\psi}(\varepsilon,t)$ is then an uniform function over the
entire FZ. In this way, we see that the system undergoes spectral ergodicity in the course of the evolution since the equipartition condition $|C_i(t)|^2\approx1/\mathcal{N}_s$
is fulfilled. 

The presence of chaos also plays an important role in the evolution at fixed force of an initial condition after a quench $F:0\rightarrow F_r$. This latter
is straightforwardly shown by computing the long-time average of the basis projector $\hat P_{\gamma}\equiv|\gamma\rangle\langle \gamma|$, i.e. 
\begin{equation}\label{eq:21}
\overline{\langle\psi(t)|\hat{P}_{\gamma}|\psi(t)\rangle}=
\lim_{\tau\rightarrow\infty}\frac{1}{\tau}\int^{\tau}_0 dt\;\langle\psi(t)|\hat{P}_{\lambda}|\psi(t)\rangle.
\end{equation}
\noindent To compute the above average we can use the evolution operator of the Floquet formalism given in Ref.~\cite{BuchJosaB1995}:
\begin{equation}\label{eq:22}
 \hat{U}(t_2,t_1)=\sum_{j,k,k'}e^{-i\varepsilon_j(t_2-t_1)}e^{-i\omega_B kt_1}e^{i\omega_B k't_2}| \phi^{k'}_{\varepsilon_j}\rangle\langle
 \phi^{k}_{\varepsilon_j}|.
\end{equation}

Choosing the initial state to be, for example, the state $|\psi(0)\rangle=|\gamma\rangle$, and assuming non degenerancies of the Floquet eigenenergies, one finds
\begin{equation}\label{eq:23}
\bar{P}_{\gamma}(F_r;|\gamma\rangle)\approx\xi_{\gamma}=\sum_j p^{\gamma}_j \langle\varepsilon_j|\hat{P}_{\alpha}|\varepsilon_j\rangle,
\end{equation}
\noindent where $\xi_{\gamma}=\sum\nolimits_j|\langle \gamma|\varepsilon_j(F_r)\rangle|^4$ and the right hand term is just the spectral average of the projector
$\hat P_{\gamma}$ \cite{CarlosThesis2013}. Here the occupation probabilities satisfy the normalization condition $\sum_jp^{\gamma}_j=1$. The strong mixing properties of 
the spectrum, which give rise to quantum chaos, are thus also responsible for two dynamical processes: diffusion and relaxation of the system initially prepared in 
$|\psi(0)\rangle$, either by sweeping across the spectrum ($F(t)\sim\alpha t$) or by the dynamical evolution after the quench to a fixed tilt $F=F_r$. Note that 
$\bar{P}_{\gamma}(F_r;|\gamma\rangle)$ is nothing else but the long-time average of the survival probability for the initial state $|\psi(0)\rangle=|\gamma\rangle$.

A basic feature of chaos is that all possible dynamical processes take place with the same probability. The result is a mixture of all different
time scales in the evolution. For a system started at $F_0$ very far away from $F_r$, the only possible transitions are intra-manifold ones, which occur due to the hopping transitions,
i.e., by $J_{\beta}$ in Eq.~(\ref{eq:02}). In this process, for a given initial state $|\gamma\rangle$, one expects that its survival probability
$\bar P_{\gamma}$ showed collapses but also some revivals before $F_r$. Once the mixing between neighbor manifolds takes place, the system diffuses, and 
this effect is very much enhanced when crossing the AC structure. One way to characterize the manifold mixing is by defining the degree-of-mixing
parameter 
\begin{equation}\label{eq:24}
 \zeta(t)=1-\sum\nolimits_M (p_M(t))^2,\;\;p_M(t)=\langle\psi(t)|\hat P_M|\psi(t)\rangle,
\end{equation}
\noindent with $\hat P_M$ as defined in Sec.~\ref{sec:1a}. Clearly, in the case of a fully chaotic RET domain no revivals are observed, therefore 
maximal mani\-fold mixing arises. $\zeta(t)$ is thus maximized, and its maximal value is given by $\zeta_{\rm max}\approx 1-1/N$, with $N$
the total particle number. The system is subjected to a dynamical relaxation process, which is characterized by a power-law (scale-free) decay
of the localization measures. To see this we look at the time average function defined as:
\begin{equation}\label{eq:25}
 {\rm T-averaged}\;\;h(t)=\frac{1}{\Delta T_{k}}\sum^{\Delta T_{k}}_{t=0} h(t),
\end{equation}
where $h(t)$ is either the inverse participation ratio or the Shannon entropy. 

\begin{figure}[t]
\centering
\includegraphics[width=0.85\columnwidth]{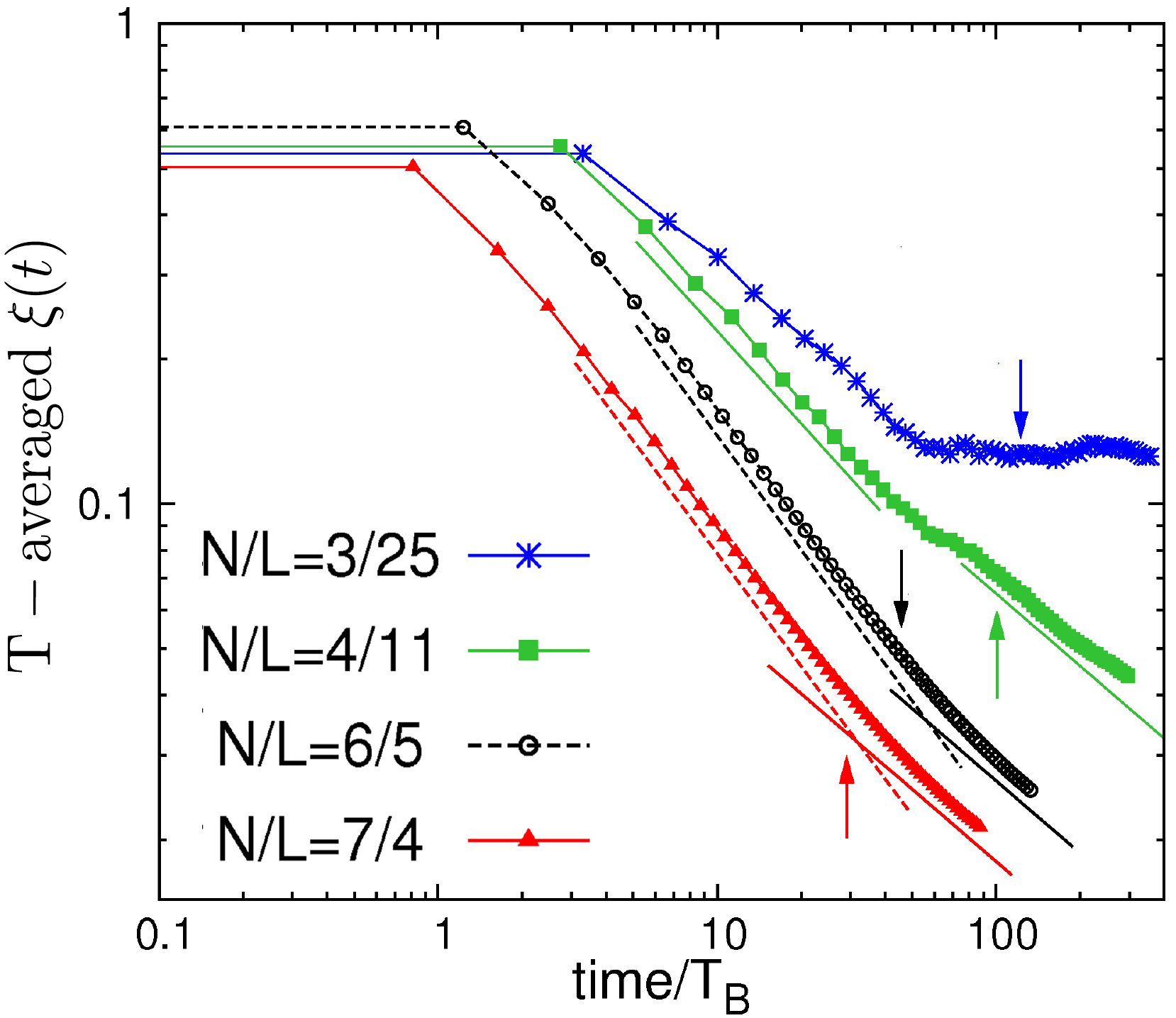}
\caption{\label{fig:7}(Color online) Relaxation: time-averaged inverse participation ration $\xi$ for 
$N/L=4/11,6/5$ and $7/4$. The straight lines indicates the power-law tendency of the diffusion processes with $t^{-0.78}$
for chaotic spectra before the resonance $F_{r=1}$ and $t^{-0.5}$ for $F>F_r$. The change of the exponents corresponds to 
a slowing down of the spreading of the evolved state since in the chaotic case the maximal delocalization has already occurred. For not fully
chaotic spectra, $N/L=4/11$, the tendency to a power-law $t^{-0.64}$ is destroyed by the emerging relocalization as explained in Sec.~\ref{sec:3b},
see c.f. $N/L=3/25$.}
\end{figure}

In Fig.~\ref{fig:7}(a) we show a double-logarithmic plot of the time evolution of the inverse participation ratio. Two different power-laws are
observed (see straight lines). In a first instance, the full chaotic spectra $N/L=6/5,7/4$ show a well defined decay $t^{-\nu}$ with exponents
$\nu\approx0.78$ before $F_r$ (indicated by the arrows). Afterwards a slowing down for $F>F_r$ occurs due to the maximization of the spreading in the
instantaneous eigenbasis. The exponent for this region is $\nu\approx0.5$. In the latter case the straight lines depict the tendency of the evolution 
if $\Delta T$ is extended to reach again the off resonant regime where the mixing is suppressed. In this regime the equidistribution of the probability
over the energy space remains unchanged, explaining the slowing down in Fig.~\ref{fig:7}.

For $N/L=3/25,4/11$ deviations from chaotic spectra occur (see Sec.~\ref{sec:2c}). Here the time-evolved states undergo different processes in the course of the evolution. We observe a tendency to a power-law with $\nu\approx 0.64$ ($N/L=4/11$), but also a slowing down before $F_r$. Interestingly, right before
$F_r$ localization occurs. After $F_r$, the short time decay presents also a power-law exponent $\nu\approx 0.5$ which implies a certain stabilization (slowing
down). Yet for long times, one observes final relocalization highlighted by a second slowing down. This is most clearly seen for $N/L=3/25$, where the decay stops completely 
(see stars Fig.~\ref{fig:5}(a)). It does, however, not undergo any equilibration. We see that chaos, apart from generating
strong band mixing, also induces a fast decay to the equilibrium values set by the GUE limits of Eq.~(\ref{eq:20}). In Fig.~\ref{fig:7} one can also notice that the 
system diffuses the slower, the smaller the filling factor $N/L$ is. 

In our case, the equilibrium is defined in the context of the energy shell approach \cite{LFSantosPRL2012}. However, the connection is not straight forward,
since in our case the distribution of coefficients $C_i$ as a function of the energies within the Floquet zone is nearly a flat function. Therefore, the
LDOS is an extended function over the entire spectrum. In the energy shell approach, the distribution of the coefficients is expected to be gaussian-distributed. To 
solve such a discrepancy, one must do an unfolding of the distribution $P_{\psi}$, or equivalently one can fold the gaussian profile into the Floquet zone (FZ).
The latter method is straight forward since the resulting function is a {\it normal wrapped distribution} \cite{Fisher1996}, which is a
periodic function in the energy domain, meaning in our case, in the FZ zone.

\begin{figure}[t]
\centering
\includegraphics[width=\columnwidth]{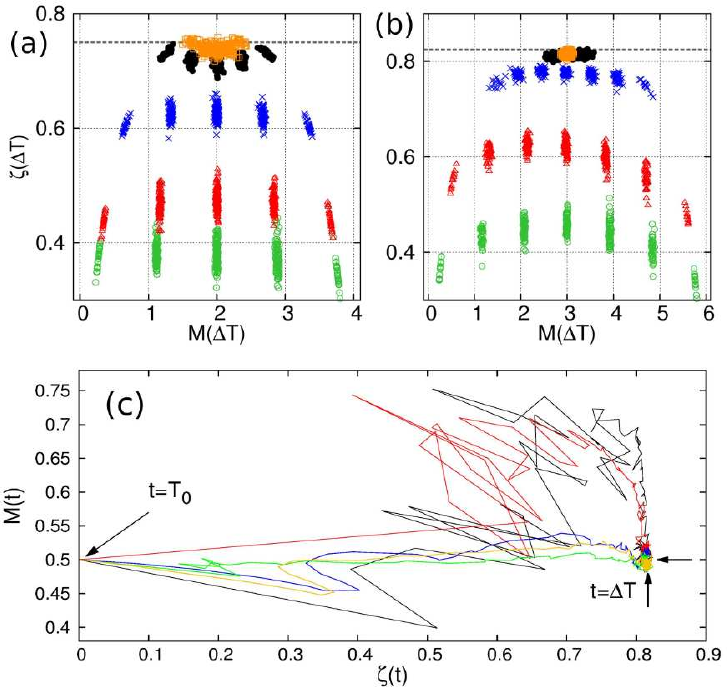}
\caption{\label{fig:8}(Color online) Manifold mixing: (a-b) depict the final time manifold degree of mixing $\zeta(\Delta T)$ and manifold creation 
$M(\Delta T)$ for different band gaps: $\Delta_g=2.53$ (green \textopenbullet), $\Delta_g=1.16$ (red $\vartriangle$), $\Delta_g=0.556$ 
(blue $\times$), $\Delta_g=0.285$ (black \textbullet) and $\Delta_g=0.155$ (dark orange $\square$). To this end, more than $\mathcal{N}_s/2$ initial 
states were evolved starting at $F_0=0$, for (a) $N/L=4/11$ and (b) $N/L=6/5$. Panel (c) depicts $\zeta(t)$ vs $M(t)$ with $t$ as a parameter, for 
$\Delta_g=0.155$. Note that all trajectories in the plane $\zeta-M$ converge to the equilibrium point $(\zeta_{\rm max},M=N/2)$ (see main text).}
\end{figure}

As final result, we show the dynamical creation and destruction of the $M$-manifolds. This is done by computing the manifold mixing degree $\zeta(\Delta T)$
(or the localization measures) and the manifold number $M(\Delta T)$ at the final time, i.e., at $F_f>F_r$. The results are shown in Fig.~\ref{fig:8} for 
(a) $N/L=4/11$ and (b) $N/L=6/5$. For this calculation we have evolved more than $\mathcal{N}_s/2$ different initial conditions belonging to all possible sets
of manifolds. Some of them are plotted in Fig.~\ref{fig:8}(c), which shows the trajectories in the plane $\zeta$--$M$. One must keep in mind that the 
number of manifolds is $N+1$. Taking the final time $\Delta T$ as a parameter, the panels (a-b) show the destruction
of the manifolds as the gap decreases below a critical value $\Delta_g\approx 0.285$. Here one can no longer identify the separated
bunches of states with well defined manifold numbers. The latter dynamical effect is expected according to the discussion of Sec.~\ref{sec:1a}. In addition,
the mixing does not depend on the class of initial states. For the fully chaotic spectrum $\Delta_g=0.155$, all final states are completely delocalized. 
This implies that $M= N/2$, and $\zeta(\Delta T)\approx 1-1/N$, as around (\ref{eq:24}), dashed line in Figs.~\ref{fig:8}(a-b), which is exactly the equilibrium condition. We thus confirm 
that the outgoing state after the passage across the RET is indeed an equilibrium state for which the respective entropy is maximized due to the 
presence of fully chaotic many-body AC structure.

\section{Conclusions.}\label{sec:4}
We have studied in detail the spectral properties of a many-body two-band Wannier-Stark system with particular emphasis on the resonant tunneling regime.
Depending on the strength of the interaction, on the band gap between the two lowest energy bands, and on the filling factor,
the spectra show a regular to quantum-chaotic transition. This allows us to study the diffusive properties of generic energy spectra in Hilbert
space. The spectral characteristics can be probed by quantum sweeps of different initial conditions across the resonant regions, by using the Stark force
as a time-dependent control parameter. In this way, we can clearly distinguish the dynamics depending on the various spectral characteristics. Relaxation toward
equilibrium, corresponding to a maximal delocalization in energy (Hilbert) space occurs for quantum chaotic spectra. Interestingly, the spectral 
ergodicity arises in both types of dynamics, either by sweeping across the chaotic many-body RET regime, or by a quench with additional free evolution
at fixed tilt. 

In the case of regular or mixed spectra, showing a poissonian component in the nearest neighbor statistics, localization of the instantaneous states 
preserves in the dynamics. In this case, full ergodicity cannot arise. The manifold approach developed here, starting from the single-particle (noninteracting) case,
has proved to be a good tool to analyze the localization-delocalization transition. It provides an intuitive picture based on the mixing of the manifolds
during the temporal evolution. The transition between the various regimes can be controlled by means of the system parameters, in particular, the interparticle
interaction, the filling factor and the Stark force.

As explained in the appendix, our two-band system can readily be realized in the experiment. Its implementation is based on the miniband structure, which can be
easily engineered using a double period one-dimensional lattice \cite{WeitzPRL2007,Modugno2011}. A standard procedure for controlling, i.e., the Stark force, 
is by accelerating the lattice structure by shifting the frequencies of two counter-propagating waves that generate the optical potential 
\cite{Arimondo2011,WimbPRL2007,WeitzPRL2007,ChSalomonPRL1997}. Our system offers a high controllability of all system parameters. For instance, the interparticle interaction ($g\sim a_{\rm scatt}$)
 can be changed by Feschbach resonances \cite{IblochREP2008,InsbruckArxiv2013}. The remaining parameters can be varied by using the geometry properties
of the lattice. In this way, our engineered system and results exposed in this paper open an interesting route toward the 
realization of complex many-body systems, with immediate experimental implications on coherent control of ultracold atoms
\cite{MGreiner2011,Arimondo2011,InsbruckArxiv2013,IblochPRL2007,Bloch2013,drive}. 

\section{Acknowledgments}
S.W. acknowledges financial support from the DFG (FOR760), the Helmholtz Alliance Program EMMI (HA-216), and
the HGSFP (GSC 129/1). It is our pleasure to warmly thank F. Borgonovi, P. Schlagheck, B. Fine for lively discussions and 
T. Wellens for useful comments on our implementation of a two-band Wannier-Stark system.

\appendix
\section{Wannier functions and Bose-Hubbard coefficients}\label{appx}
\begin{figure}[b]
\centering \includegraphics{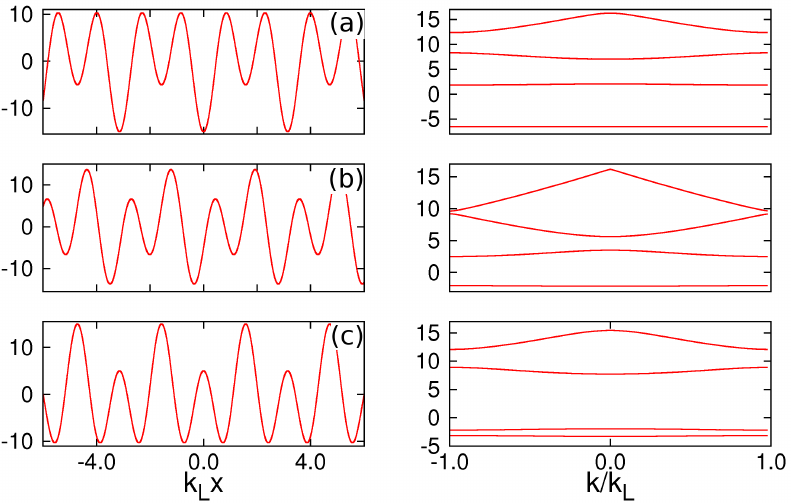}
\caption{\label{fig:apx01}(Color online): Profile of the optical lattice (left) and its respective band structure (right), for $V_0=3$ and $z_0=4$, 
as function of the phase difference $\phi$. (a) $\phi=0$, (b) $\phi=\pi/2$ and (c) $\phi=\pi$.}
\end{figure}
Here we show how the coefficients in Eq.~(\ref{eq:02}) are computed for an experimental realization with ultracold bosons in an optical lattice. 
We based our calculation on the single-particle Wannier functions, which are
localized within each site. We suggest to use a double periodic optical lattice, as experimentally realized in \cite{WeitzPRL2007,Modugno2011}:
\begin{equation}\label{eq:appx01}
V(x)=-V_0\left[\cos(2k_Lx)+z_0\cos(4k_Lx+\phi)\right], 
\end{equation}
with $k_L$ being the recoil momentum, and the recoil energy $E_r=\hbar^2k^2_L/2m_0$. The energy dispersion relation 
is computed by diagonalizing the Hamiltonian $\hat{H}_0=\hat{p}^2/2m_0+V(x)$ as function of the lattice parameters: the depth $V_0$ of the
lattice, the ratio $z_0=V_1/V_0$ between the amplitudes of the two lattices, and the phase difference $\phi$. One can thus appropriately 
engineer a periodic potential for which the respective two lowest Bloch bands are well separated from all higher energy bands \cite{WeitzPRL2007}, as
shown in Fig.~\ref{fig:apx01}. In this way, by choosing a relative phase $\phi\equiv\pi$ and appropriate values of $z_0$, we can 
work with a realistic closed two-band model, represented by our Hamiltonian in Eq.~(\ref{eq:01}). 
The Wannier functions are defined through the Fourier transform of the Bloch functions, $\psi_{\beta,{\tilde{k}}}(x)=e^{i\tilde{k}x}u_{\beta}(x)$,
in the first Brillouin zone (BZ) as
\begin{equation}
\chi_{\beta}(x)=\int_{BZ}e^{-i\tilde{k} x_l}\psi^{\beta}_{\tilde{k}}(x)d\tilde{k},
\end{equation}
with $u_{\beta}(x)=u_{\beta}(x+d_L)$. $d_L$ is the spatial periodicity of the lattice and $x_l\rightarrow d_L l$. Since $\tilde{k}=k/k_L$ is a parameter, the Bloch
functions are not unique and a phase factor can be chosen such that the Wannier functions are highly localized 
\cite{KorschPRep2002,WKohnPRB1973,TomadimPRL2008}. The latter property and the appropriate symmetry, $\chi_{\beta}(-x)=(-1)^{\beta-1}\chi_{\beta}(x)$, 
are shown to be satisfied by the following functions
\begin{eqnarray}
 \chi_{1}(x)=\frac{1}{\sqrt{N_1}}\sum_{k_{j},n}|u_n(1,\tilde{k}_j)|\cos[k_{j,n}x_0]\cos[k_{j,n}x]\nonumber\\
 \chi_{2}(x)=\frac{i}{\sqrt{N_2}}\sum_{k_{j},n}|u_n(2,\tilde{k}_j)|\sin[k_{j,n}x_0]\sin[k_{j,n}x]\nonumber,\\
\end{eqnarray}
where $k_{j,n}\equiv2n+\tilde{k}_j$. $x_0=\pm\cos^{-1}(1/4z_0)$ are the first minima position of the potential in Eq.~(\ref{eq:appx01}) around
$x=0$ and $N_{1,2}$ are normalization constants. The coefficients $u_n(\beta,\tilde{k}_j)$ are the Fourier components of the periodic 
function $u_{\beta}(x)$ given by $u_{\beta}(x)=\sum_{n} u_{n}(\beta,\tilde{k})e^{-i nx}$.
\begin{figure}[t]
\centering \includegraphics[width=\columnwidth]{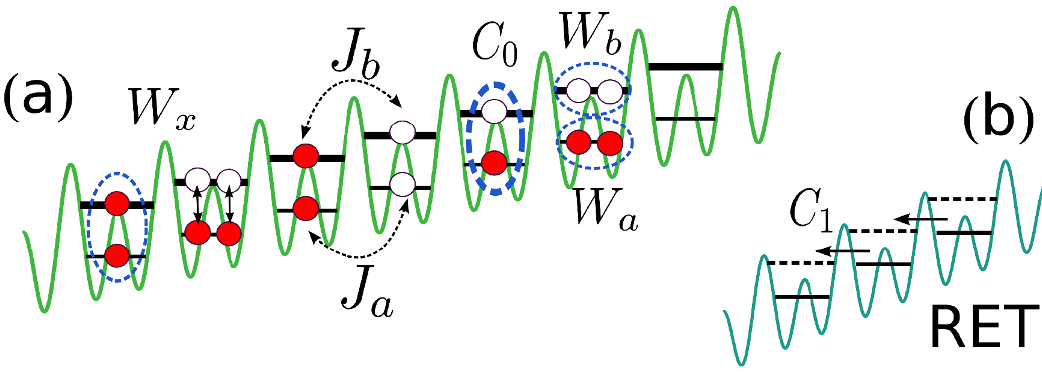}
\caption{\label{fig:apx02}(Color online): (a) Many-body processes of the two-band Bose-Hubbard Hamiltonian for a bichromatic tilted 
optical lattice. (b) {\bf R}esonant {\bf E}nhanced {\bf T}unneling (RET) condition for the nearest neighboring double wells, i.e, for a first 
order resonance.}
\end{figure}
The Bose-Hubbard coefficients, sketched in Fig.~\ref{fig:apx02}(a), are then obtained from the following relations: the hopping amplitudes $J_{\beta}$ are  
\begin{equation}
 J^{\beta}_{l-l'}\equiv \int \chi_{\beta}^*(x-x_l)H_0(x)\chi_{\beta}(x-x_{l'})\;dx=\epsilon^{\beta}_{l-l'},
\end{equation}
where $J_a\equiv J^{\beta=a}_{1}$, $J_b\equiv J^{\beta=b}_{1}$, and $\Delta_g=|\epsilon^b-\epsilon^a|=|J^{\beta=b}_{0}-J^{\beta=a}_{0}|$.
The dipole-like coupling strengths are 
\begin{equation}
C^{\beta\beta'}_{l-l'}\equiv \int \chi^*_{\beta}(x-x_l)x\chi_{\beta'}(x-x_{l'})\;dx,
\end{equation}
with $C_\mu \equiv C^{ab}_{\mu}$. Because of the high localization of the Wannier functions, coefficient with $|\mu|>0$ are at least one order of
magnitude smaller than $C_{0}$. Thus we only take into account the strength with $|\mu|=0,1$ and $2$ for the first two resonances $r=1,2$. Finally, the repulsive, intraband,
on-site interparticle interaction terms are given by
\begin{eqnarray}
W_{\beta}&\equiv& g_{1D}\int |\chi_{\beta}(x)|^4dx.\;\;
\end{eqnarray}
The interband on-site interparticle interaction is
\begin{eqnarray}
W_{x}&\equiv& g_{1D}\int |\chi_{a}(x)|^2|\chi_{b}(x)|^2dx,
\end{eqnarray}
where the interaction strength is defined by $g_{1D}=4\pi a_{1D}/m_0$, with $a_{1D}$ the one dimensional scattering constant and $m_0$ the
mass of the atoms \cite{IblochREP2008}.


\end{document}